\documentclass{PoS}

\title{Quark and Gluon Jet Energy Loss}

\ShortTitle{Quark and Gluon Jet Energy Loss}

\author{Liliana Apolin\'ario$^{a,b}$, \speaker{Jo\~ao Barata}$^c$
and Guilherme Milhano$^{a,b}$\\
\llap{$^a$}LIP, Av. Prof. Gama Pinto, 2, P-1649-003 Lisboa, Portugal\\
\llap{$^b$}Instituto Superior T\'ecnico (IST), Universidade de Lisboa, Av. Rovisco Pais 1, 1049-001, Lisboa, Portugal\\
\llap{$^c$}Instituto Galego de F\'isica de Altas Enerx\'ias (IGFAE), Universidade de Santiago de Compostela, E-15782 Galicia, Spain\\
E-mail: \email{liliana@lip.pt}, \email{joao.barata@cern.ch},
\email{gmilhano@lip.pt}}

\abstract{
We present a study of the relative energy loss of quark and gluon jets that traverse a weakly-coupled medium. The study is carried out with the jet quenching Monte Carlo event generator JEWEL which has been validated for a large set of jet observables. We find that the relative energy loss of quark and gluon jets approaches a constant value at large jet energy and that value is significantly larger (closer to 1) than the single parton expectation of $C_F/C_A$. Within JEWEL quark and gluon jets lose energy similarly.
}

\FullConference{International Conference on Hard and Electromagnetic Probes of High-Energy Nuclear Collisions\\
		30 September - 5 October 2018\\
		Aix-Les-Bains, Savoie, France}

\begin{document}

\section{Introduction}
In perturbative QCD it is well established that the energy loss rate $\frac{dE}{dx}$ for an energetic parton due to interactions with a weakly-coupled medium  is controlled by the Casimir of the corresponding representation \cite{rev}. In fact one can write \cite{hyb}: 
\begin{equation}
\frac{\left(\frac{dE}{dx}\right)^{quark}}{\left(\frac{dE}{dx}\right)^{gluon}} \propto \frac{C_F}{C_A}\stackrel{\scriptscriptstyle QCD}{=} 0.444\ldots  
\end{equation}
where $C_F=(N^2-1)/(2N)$, $C_A=N$ and $N=3$ is the number of colours in QCD, as usual. We will refer to this feature of energy loss troughout the text as Casimir scaling. A similar scaling is also present in non-perturbative approaches at the parton level \cite{hyb}.\par 
Although this scaling is well understood for a single parton, it is not known which scaling jets should obey. This a relevant issue because jets are the physical objects one can study and the transition from one parton to a multi parton state is non-trivial.\par
\begin{figure}
 \centering
\includegraphics[width=0.8\textwidth]{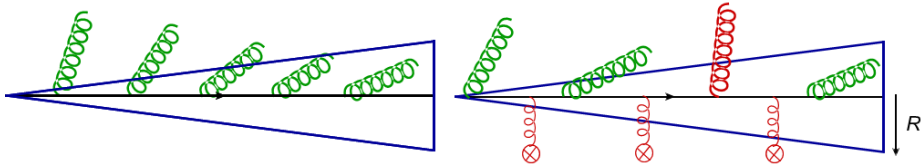}
\caption{Simplified depiction of jet evolution in vacuum (left) and in the medium (right). In both figures the leading parton is represented by a solid black line around which we draw a jet with parameter $R$. In the left plot, the only emissions (in green) are controlled by the Sudakov form factor, while the right plot also includes medium-induced radiation (in red, with scattering centres denoted $\otimes$).}
\label{fig:pic}
\end{figure}

 
Figure \ref{fig:pic} provides a simplified depiction of a jet in vacuum (left) and in the medium (right). In both figures we notice that due to the finite size of the jet, unlike the leading parton (i.e. the single parton picture) some emissions are recovered. This suggests that Casimir scaling should be broken for medium induced radiation and there should also be a scaling just due to recovering Sudakov emissions in vacuum. In this context, we expect to find a different scaling for jets in medium. 
Two factors are expected to contribute to the breaking of Casimir scaling: that since jets are extended objects, not all emissions contribute to the energy loss spectrum, and that most jet constituents, in both quark and gluon jets, are gluons. \par 

There are then two questions we will address. First, we study how the energy loss scales with the choice for the jet parameter $R$ which influences both in-vacuum and in-medium physics. The second point (and the  main motivation for our study) is to understand how medium induced radiation scales with respect to colour factors. In particular, as is mentioned and discussed above, we are interested in understanding the quark to gluon scaling.\par 
The scaling for jets has been previously addressed in \cite{S2,S1}, where the scaling is allowed to differ by a constant factor from the single parton result and in \cite{Yacine} the scaling is broken if the particle-medium interactions are relevant. In this work, we will carry a full Monte Carlo (MC) analysis to evaluate the validity of such parametrizations.
\section{Analysis details}
We generate JEWEL events \cite{JEWEL}  separately for $Z$ + quark-initiated jets and $Z$ + gluon-initiated jets. This allows to compute the flavour ratio in a controlled way. The initial parton $p_T$ is given by the proxy (back-to-back) $p_T^Z$, thus allowing to compute the energy loss ratios. We produce samples with $10^6$ events in vacuum and in the medium for a centrality $0-10  \%$, with JEWEL's default parameters. We have no recoils, so that the observables are not contaminated by the medium and we work at the parton level (no hadronization). Additionally we ignore initial state radiation (ISR) and the centre of mass energy matches the one of the latest LHC run $\sqrt{s_{NN}}=5.02$~TeV. 
\par 
We have the following selection cuts:
\begin{itemize}
    \item Z cuts: $p_T^Z>50$~GeV, $|\eta^Z|<2.5$, di-muon channel
    \item jets cuts: $p_T^{jets}>20 $~GeV, $|\eta^{jets}|<2.5$
\end{itemize}
We choose the most energetic anti-$k_t$ jet that satisfies $5\pi/6<|\phi_Z-\phi_{jet}|<7\pi/6$ (it is back-to-back with the $Z$). We check that for vacuum this selects the highest $p_T$ in more than $99\%$ of the selected events for all the cones in vacuum and in the medium we get a value of the same order. \par 
A problem which can arise is the fact that an initial hard splitting might change the flavour of the reconstructed leading jet. In JEWEL, next-to-leading-order (NLO) matrix elements are not included which already minimises the flavour changing events. In addition, kinematics and the LO QCD splittings functions further constraint this issue. A future analysis will deal with this question in a more detailed manner.
\section{Vacuum Scaling}
We first focus our study on the scaling in vacuum due to the jet definition. For that, we study the variable $(p_T^Z-p_T^{jet})/p_T^Z = \Delta p_T/p_T^Z$. This captures the total energy imbalance in the event with respect to the initial energy of the leading parton. \par
\begin{figure}[h!]
   \centering
\includegraphics[width=0.6\textwidth]{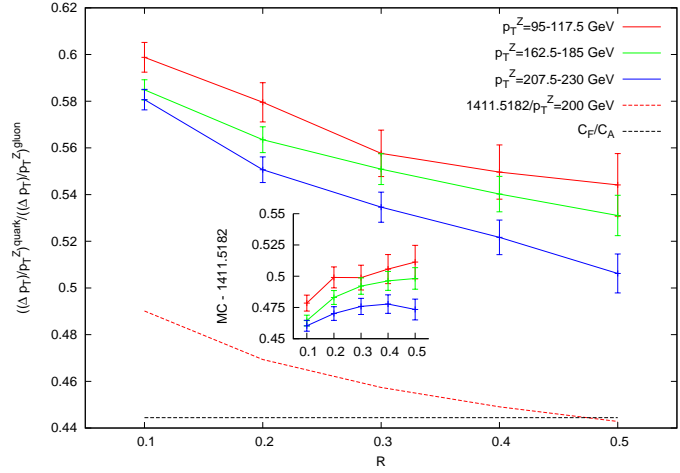}
\caption{Quark to gluon ratio scaling in vacuum as a function of the jet parameter $R$. We see that even in vacuum there exists a non trivial scaling. Our solution has a qualitative behaviour similar to the one observed in the analytic approach \cite{Salam}. In particular, it seems that the results just differ by an overall constant, which might be a consequence of using the running coupling at one-loop. Other effects implemented in the JEWEL might also contribute to this difference. In the subplot one can check that the difference between the analytic and MC solutions is roughly constant. The analytic solution was plotted using $p_T^Z$ as the relevant scale for the running coupling.} 
\label{fig:2}
\end{figure}

 
In Figure \ref{fig:2} we present the results of our analysis for 3 bins in $p_T^Z$ for the evolution of ratio of the above variable with respect to the jet parameter $R$. Additionally we present the quark to gluon ratio computed from \cite{Salam}. This solution resums to all orders terms enhanced by $\alpha_s\ln{\left(R^2\right)}$ factors, i.e. it takes into account the quark to gluon scaling for jets just due to $R$. At leading order the solution is roughly $C_F/C_A$. However, the plotted curve makes use of the solution for the running coupling at one-loop, which implies that one is free to set the overall relevant momentum scale. Therefore, we can only qualitatively compare the analytic and MC solutions. While the shapes do not seem to be in agreement, the values at large $R$ seem to match between the MC and analytic results. In any case we expect some deviation at small $R$ due to $\alpha_s\ln^2{\left(R^2\right)}$ corrections.

\section{The Breaking of Casimir Scaling}
\begin{figure}[h!]
 \centering
\includegraphics[width=0.65\textwidth]{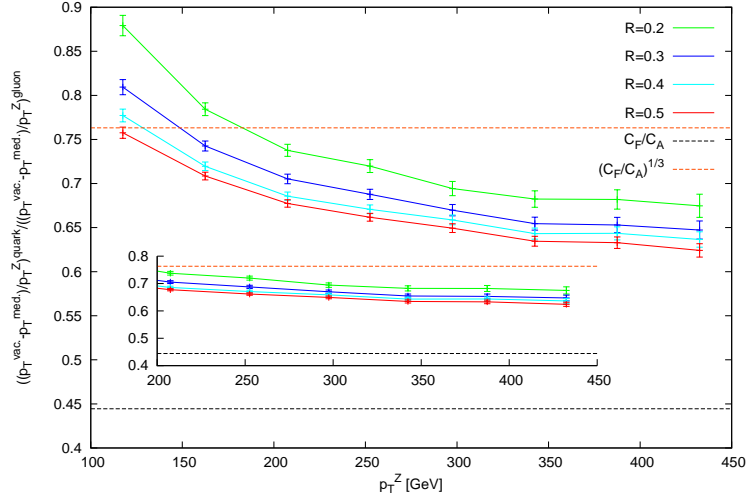}
\caption{Evolution of the quark to gluon ratio for the energy lost just due to interactions with the medium for 4 values of $R$. The orange dashed line represents the scaling used in \cite{hyb}.} 
\label{fig:3}
\end{figure}

The previous section showed that even without medium effects, the jet definition leads to a non-Casimir quark to gluon scaling due to the finite size of the jet. We are now interested in the medium effects, thus we must define a suitable variable that just captures this aspect. For that we consider $(\Delta p_\perp^{medium}-\Delta p_\perp^{vacuum})/p_\perp^Z=(p_T^{vacuum}-p_T^{medium})^{jet}/p_\perp^Z$ which captures the energy imbalance due to the medium interaction with respect to the initial parton energy, when averaged over all events. In figure \ref{fig:3} we present our results for the evolution with $p_T^Z$ for four different values of $R$. It is possible to see that Casimir scaling is explicitly broken for jets. In addition for high enough $p_T$, we recover a constant scaling for all $R$, which lies between the hybrid model and single parton pictures. We computed the average value $\bar{x}_R $ in the domain $200<p_T^Z<500$~GeV and obtained $\bar{x}_{0.2}=0.732 \pm 0.027$, $\bar{x}_{0.3}=0.696 \pm 0.021$, $\bar{x}_{0.4}=0.679 \pm 0.018$ and $\bar{x}_{0.5}=0.668 \pm 0.016$. 

\section{Conclusions}
The relative energy losses of quark and gluon jets in the vacuum and in a QCD medium have been studied. For the vacuum evolution we see that our results are qualitatively in agreement with the solution presented in \cite{Salam}. This seems to show that at this level there are no additional physics effects besides the jet radius definition, thus preventing additional bias in the analysis. \par 
In the case of in-medium evolution, our results show that the Casimir scaling is clearly broken for a wide range in $p_T^Z$. Additionally, for enough energetic jets the scaling is broken by an overall constant, which has a very small dependence on $R$. Interestingly the quark to gluon ratio yields an average value much closer to unity than $C_F/C_A$. This result seems to indicate that in the medium, quark and gluon jets are much more similar than what could be naively expected from single parton computations. We interpret this fact as a consequence that a quark jet in the medium will be highly populated with gluons, thus leading to a "washing out" of the initial flavour. \par 
Our result is qualitatively in agreement with the parametrization used in \cite{S2,S1}, although the results can not be quantitatively compared. Additionally, we would like to point out that the motivation presented in \cite{S2,S1} goes against our conclusions. A more in depth discussion on the results and further details will be presented in a forthcoming paper.

 \section*{Acknowledgements}
The authors acknowledge financial support from Funda\c{c}\~ao para a Ci\^encia e Tecnologia (FCT-Portugal) under contracts CERN/FIS-PAR/022/2017.

\end{document}